\documentclass[fleqn,11pt]{article}
\usepackage{amsfonts,amssymb,cite}
\usepackage{graphicx}

\def\noi{\noindent}

\newcommand{\Title}[1]{\noi {{\Large\bf #1}}\\[1ex]}

\newcommand{\Author}[2]{\noi{\bf #1}\\[2ex]\noi{\normalsize\it #2}\\}

\newcommand{\Abstract}[1]{\vskip 2mm \begin{center}
        \parbox{16.4cm}{\small\noi #1} \end{center}\medskip}
\newcommand{\foom}[1]{\protect\footnotemark[#1]}

\def\nqq{\hspace*{-2em}}

\def\Jl#1#2{#1 {\bf #2},\ }

\def\ApJ#1 {\Jl{Astroph. J.}{#1}}
\def\CQG#1 {\Jl{Class. Quantum Grav.}{#1}}
\def\DAN#1 {\Jl{Dokl. AN SSSR}{#1}}
\def\GC#1 {\Jl{Grav. Cosmol.}{#1}}
\def\GRG#1 {\Jl{Gen. Rel. Grav.}{#1}}
\def\JETF#1 {\Jl{Zh. Eksp. Teor. Fiz.}{#1}}
\def\JETP#1 {\Jl{Sov. Phys. JETP}{#1}}
\def\JHEP#1 {\Jl{JHEP}{#1}}
\def\JMP#1 {\Jl{J. Math. Phys.}{#1}}
\def\NPB#1 {\Jl{Nucl. Phys. B}{#1}}
\def\NP#1 {\Jl{Nucl. Phys.}{#1}}
\def\PLA#1 {\Jl{Phys. Lett. A}{#1}}
\def\PLB#1 {\Jl{Phys. Lett. B}{#1}}
\def\PRD#1 {\Jl{Phys. Rev. D}{#1}}
\def\PRL#1 {\Jl{Phys. Rev. Lett.}{#1}}

\def\lal{&&\nqq {}}

\def\beq{\begin{equation}}
\def\eeq{\end{equation}}
\def\bear{\begin{eqnarray}}
\def\bearr{\begin{eqnarray} \lal}
\def\ear{\end{eqnarray}}
\def\earn{\nonumber \end{eqnarray}}

\usepackage{amsmath,amssymb}
\usepackage{xcolor}
\usepackage{verbatim}

\newtheorem{Mark}{Comment}
\newcommand{\Marking}[2]{%
\begin{Mark}\hspace{-4pt}\textbf{. (#1)}\\
#2
\end{Mark}}
\newcommand{\Fig}[3]{%
\begin{center}
\parbox{8cm}{%
\refstepcounter{figure}\includegraphics[width=8cm,height=#2cm]{#1} \noindent Fig. \thefigure:\quad
#3}\end{center}}
\newcounter{strochka}

\newcounter{spisok}
\begin{document}
\thispagestyle{empty}
\twocolumn[

\vspace{1cm}

\Title{Formation of supermassive nuclei of Black holes in the early\\[8pt] Universe by the mechanism of scalar-gravitational instability.\\[8pt] III. Large scale picture.   \foom 1}

\Author{Yu. G. Ignat'ev}
    {Institute of Physics, Kazan Federal University, Kremlyovskaya str., 16A, Kazan, 420008, Russia}


\Abstract
 {The dependence of the parameters of the evolution of scalarly charged Black Holes (BHs) in the early Universe on the parameters of field-theoretic theories of interaction, the influence of the geometric factor of the structure of the relative position of BHs on the limitation of their maximum mass are studied, the problem of the metric of a scalarly charged BH in a medium of expanding scalarly charged matter is discussed, the expression is obtained for the macroscopic cosmological constant at late stages of expansion, generated by quadratic fluctuations of the metric, connecting the value of the cosmological constant with the BH masses and their concentration.
  \\[8pt]
}
\bigskip

] 

\section*{Introduction}
In the first part of the author's article \cite{Yu_GC_23_No4} two problems were formulated that need to be solved in the theory of the formation of supermassive black hole (SSBH) nuclei in the early Universe using the mechanism of scalar-gravitational instability of the cosmological medium of scalar-charged fermions, in order to lead it in accordance with the observed picture:\\
1. According to the results presented, the process of increasing the SSBH mass does not stop when the required mass \eqref{M_nc} is reached (see \cite{SMBH1e} -- \cite{SMBH2e})
\begin{equation}\label{M_nc}
m_{ssbh}\sim 10^4\div 10^6 M_\odot\approx 10^{42}\div10^{44}m_{\mathrm{pl}},
\end{equation}
but continues endlessly. Now we need to find a mechanism to stop this process.\\
2. What does the large-scale structure of the Universe become after the completion of this process, what is the fate of the matter that fell into the sphere of influence of SSBH?

The first question was partially answered in the second part of the article \cite{Yu_GC_24_No1} - the process of evolution of \emph{spherical} scalar-gravitational disturbances ends quite quickly automatically precisely due to the geometric factor of spherical symmetry. As shown in this work, the process of SSBH formation is determined by the fundamental parameters of the cosmological model:
\[\mathbf{P}= [[\alpha,m_s,e,\pi_0],\Lambda],\]
where $\alpha$ is the self-interaction constant of the Higgs potential, $m_s$ is the mass of scalar bosons, $e$ is the scalar charge of fermions, $\pi_0$ is their initial Fermi momentum, $\Lambda$ is the cosmological constant, as well as initial conditions, which in the simplest case (if the initial values of the derivatives of functions are equal to zero) can be written in three quantities
\[I=[\Phi_0,m_0,q_0],\]
where $\Phi_0$, $m_0$, $q_0$ are the initial values of the scalar potential, central singular mass and perturbation charge, respectively. In this article, we, firstly, specify the dependence of the parameters of the evolution of BHs in the early Universe on the parameters of the field theoretical model of interactions, secondly, we study large-scale geometric factors that stop the process of growth of the BH mass, and thirdly, we consider a possible large-scale picture of the Universe at the end of SSBH formation process.

\section{Formation and evolution of black holes in various field-theoretical models}
In \cite{Yu_GC_24_No1} three \emph{similar} process models are considered:\\
basic model corresponding to the Planck interaction scales\footnote{We use the Planck system of units $G=c=\hbar=1$.}, --
\begin{equation}\label{Par}
\mathbf{P_0} =\bigl[\bigl[1,1,1,0.1\bigr],3\cdot10^{-6}\bigr];
\end{equation}
field theoretical model SU(5), resulting from the base model with similarity coefficient $k=10^{-5}$:
\begin{eqnarray}\label{ParSU5}
\mathbf{P_{SU(5)}} =\bigl[\bigl[10^{-10},10^{-5},\sqrt{10}\cdot10^{-3},\nonumber\\
\;\;\;\sqrt{10}\cdot10^{-4}\bigr],3\cdot10^{-16}\bigr],\quad (k=10^{-5}
\end{eqnarray}
and the standard field theoretical model SM - with similarity coefficient $k=10^{-15}$:
\begin{eqnarray}\label{ParSM}
\mathbf{P_{SM}} =\bigl[\bigl[10^{-30},10^{-15},\sqrt{10}\cdot10^{-8},\nonumber\\
\;\;\;\sqrt{10}\cdot10^{-9}\bigr],3\cdot10^{-36}\bigr],\quad (k=10^{-15}).
\end{eqnarray}
In this case, as shown in \cite{Yu_GC_24_No1}, the final parameters of the process significantly depend on the initial conditions only by the factor of the location of the initial state of the system in relation to \emph{unstable} in this model,
which corresponds to the value of the scalar potential $\Phi_+=1$ in the models under consideration. Namely, we will denote process models with an initial state ``above'' stable $\Phi(0)>\Phi_+$ by the upper symbol $^+$, and models with an initial state ``below'' stable $\Phi(0) <\Phi_+$ with the top symbol $^-$: $M^+_0,M^-_0$, etc.

Taking into account the properties of the similarity transformation (see \cite{Yu_GC_24_No1}), we present approximate results of numerical simulation of the evolution of the BH mass for three similar cosmological models.

\begin{flushleft}
\refstepcounter{table}Table \thetable. \label{tab1}Maximum mass of black hole\\[12pt]
\begin{tabular}{|l||c|c|c|c|c|}
\hline
& & & & &\\[-2pt]
M & $m_{max}$ & $m^\odot_{max}$ & $t_{max}$ & $t_{su(5)}$ & $t_{sm}$ \\[6pt]
\hline
& & & & & \\[-10pt]
$\mathbf{M^-}$ & $10^{27}$ & $10^{-11}$ & $2.5(2)$ & $2.5(7)$ & $2.5(17)$ \\[6pt]
$\mathbf{M^+}$ & $10^{52}$ & $10^{14}$ & $3.7(2)$ & $3.7(7)$ & $3.7(17)$ \\[6pt]
\hline
\end{tabular}
\end{flushleft}
\hrule
\vspace{6pt}
Explanations for Table \ref{tab1}: $t_{max}$ -- time at which the maximum mass $m_{max}$ of the black hole is reached, for the model \eqref{Par}, $t_{su(5)}$ -- for model $\mathbf{P_{SU(5)}}$ and $t_{sm}$ -- for model $\mathbf{P_{SM}}$; $m^\odot_{max}$ -- the value of this mass in solar mass units; numbers in parentheses indicate order.

Let us note, firstly, that in all cases the initial singular mass in the spherical perturbation was assumed to be equal to $m(0)= m_{Pl}$ (!). When the initial mass increases by $p$ times, it is necessary to increase the maximum mass by $p$ times.
From Table \ref{tab1}, secondly, one can see that models of the $\mathbf{M^-}$ type lead to maximum masses of formed BHs that are approximately 25(!) orders of magnitude smaller than models of the $\mathbf{M type ^+}$. Thirdly, the smaller the fundamental constants $\mathbf{P}$, the longer the BH formation time. Fourth, and finally, the maximum masses of formed BHs in all models of the $\mathbf{M^+}$ type that suit us, even in the $\mathbf{M^+_0}$ model, are many orders of magnitude higher than the required SSBH masses \eqref{M_nc}. Therefore, although the problem with limiting the mass of SSBH was fundamentally resolved in the work \cite{Yu_GC_24_No1}, the answer to the question why the masses of SSBH are limited precisely by the limit \eqref{M_nc} was not obtained in this work.

In this paper we will try, firstly, to answer this question, as well as the second question, which, as it turns out, is closely related to the first.
\Fig{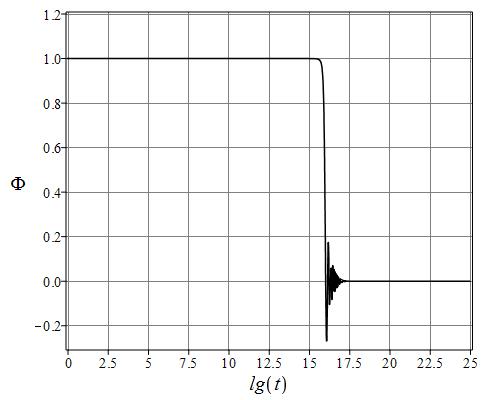}{6}{\label{Ignatev1}Cosmological evolution of the scalar field potential $\Phi(t)$ in the $\mathbf{SM^+}$ model \eqref{ParSM}.}

In what follows, we will consider the standard interaction model $\mathbf{SM}$ \eqref{ParSM} as the base model. On Fig. \ref{Ignatev1} -- \ref{Ignatev2} shows an example for the standard model of the evolution of cosmological parameters.
\Fig{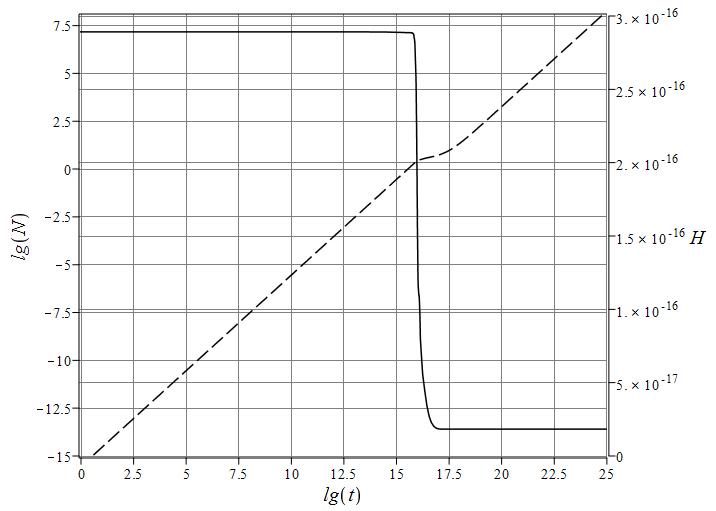}{6}{\label{Ignatev2}Dependence of the number of e-folds $N=\ln{a(t)}$ on cosmological time $t$ (dashed line) and the Hubble parameter (solid line) in the $\mathbf{SM^+}$ \eqref{ParSM}.} model

From these graphs it is clear, firstly, that at the moment of time ($t\sim 10^{16}$) there is a sharp transition of the cosmological system from a state with a scalar potential $\Phi_+\approx1$ to a state with $\Phi_\infty \to 0$, accompanied by microscopic oscillations of the scalar potential (Fig. \ref{Ignatev2}). Secondly, this transition is accompanied by an equally rapid decrease in the Hubble parameter: from $H_1\approx 2.8888\cdot 10^{-16}$ to $H_2\approx 1.8300\cdot 10^{-17}$ - while the rate of inflation expansion decreases by approximately 16 times, which corresponds to a decrease in the effective cosmological constant $\Lambda$ by approximately 250(!) times (Fig. \ref{Ignatev3}). As shown in the previous parts of the work, these transitions arise due to the instability of the cosmological system in the state \textbf{1}, as a result of which the cosmological system passes into a stable state \textbf{2}, characterized by a lower inflation rate. Thirdly, according to the results of the previous article \cite{Yu_GC_24_No1}, it is precisely during this transition that the rapid growth of the BH mass begins.

On Fig. \ref{Ignatev3} shows the results of numerical simulation of the cosmological evolution of the mass of a black hole in spherical perturbations in comparison with the formula for the mass in the $n$ harmonic of a plane perturbation, based on qualitative considerations (see \cite{Yu_GC_23_No4})
\begin{equation}\label{M_n-H0}
m(n,t)=\frac{4\pi}{n^3}H^2_0\mathrm{e}^{3H_0 t}.
\end{equation}
The results of numerical modeling confirm the approximate results of Table \ref{tab1}: a noticeable increase in mass begins immediately after the transition of the cosmological system to a stable state (see Fig. \ref{Ignatev2}) at time $t\backsimeq 4\cdot10^{16} $, the mass reaches its maximum value $m_{max}\approx 10^{52}$ at time $t_{max}\approx 4\cdot10^{17}$. After reaching the maximum, the mass begins to fall and reaches a constant limit of $m_\infty\approx 10^{42}$. Let us immediately note that the drop in the BH mass is a drawback of the linear approximation of the perturbation theory used. The mass of a macroscopic black hole, as is known, can only increase with time.
In the linear approximation of perturbation theory, the geometric properties of the black hole horizon do not appear. In this regard, the maximum mass of the black hole obtained within the framework of linear perturbation theory should be taken as the final mass of the formed black hole.
\Fig{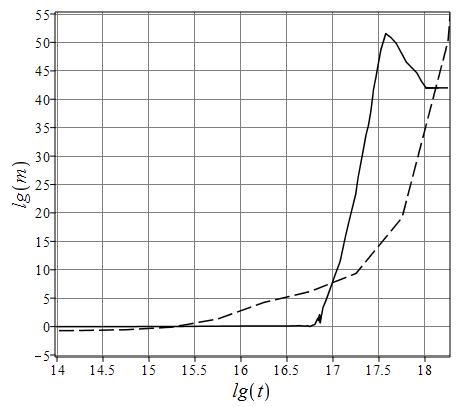}{6}{\label{Ignatev3}Cosmological evolution of the BH mass in the model of spherical perturbations (solid line) and according to the qualitative formula (dashed line) in the interaction model $\mathbf{SM^+}$ \eqref {ParSM}.}
\Marking{About estimates}{\label{mark1}Next, comparing the results of numerical modeling with the results obtained on the basis of the qualitative formula \eqref{M_n-H0}, we see that the latter give values for the mass of the black hole, achieved at the moment $t_{ max}$ is 10 orders of magnitude smaller than the exact results of numerical simulation. Or, in other words, the value $m_{max}$ is achieved based on the qualitative formula \eqref{M_n-H0} at times 2 orders of magnitude later than $t_{max}$. This must be kept in mind in the future when assessing the cosmological process of SSBH formation.}

\section{End of the SSBH formation process\label{razdel2}}
In the previous parts of the article, we considered the evolution of a single local spherical disturbance in the Friedmann Universe
\begin{equation}
ds^2=dt^2-a^2(t)(dx^2+dy^2+dz^2),
\end{equation}
$dx^2+dy^2+dz^2\equiv d\mathbf{r}^2$. In fact, such disturbances are formed, albeit randomly, but taking into account the macroscopic homogeneity and isotropy of the Universe, on average, uniformly. Of course, the initial parameters of these disturbances are largely random, but nevertheless, to simplify the model, we will assume them to be the same. Let $\tau_g$ be the moment in time of the birth of black holes (BHs) in scalar-gravitational disturbances. At the inflation stage\footnote{where $H_0=H_1$ or $H_0=H_2$ depending on the stage of evolution.}
\begin{equation}\label{inflat}
a(t)=\mathrm{e}^{H_0t}
\end{equation}
according to \cite{Yu_GC_23_No4} this moment in time is equal to
\begin{equation}\label{tau_g0}
\tau_g\backsimeq\frac{1}{H_0}\ln\biggl(\frac{n}{\sqrt{8\pi}H_0}\biggr),
\end{equation}
where $n\equiv|\mathbf{n}|$ is the wave number of the perturbation mode $\exp(i\mathbf{nr})$. The mass $m_g$ of a newborn black hole at the inflation stage does not depend on the wave number $n$ \cite{Yu_GC_23_No4}
\begin{equation}\label{M_g}
m_g\backsimeq\frac{1}{4\sqrt{2\pi}}\frac{1}{H_0}.
\end{equation}
Let us take into account that the disturbance wave\-length $\lambda(t)$ is related to the wave $n$ relation \cite{Land_Field}
\begin{equation}\label{l,n}
\lambda(t)=\frac{a(t)}{n}\Rightarrow n=\frac{a(t)}{\lambda(t)}\equiv \frac{a(\tau_g)}{\lambda (\tau_g)}
\end{equation}
And
\begin{equation}\label{lambda_g}
\lambda(\tau_g)\gtrsim 2m_g\Rightarrow n\backsimeq \sqrt{8\pi}H_0.
\end{equation}
Then we obtain an estimate for the time of BH birth in unstable disturbance modes:
\begin{equation}\label{tau_g}
\tau_g\gtrsim\frac{1}{H_0}\ln 2.
\end{equation}
Thus, we get for example in Fig. \ref{Ignatev3}:
\begin{eqnarray}
\mathbf{1:} & \tau_g\backsimeq 2.4\cdot 10^{15}; & m_g:\backsimeq 3.5\cdot 10^{14},\\
\mathbf{2:} & \tau_g\backsimeq 3.7\cdot 10^{17}; & m_g:\backsimeq 5.3\cdot 10^{15}.
\end{eqnarray}

Let further $\nu(t)$ be the average \emph{number density} of identical black holes formed at $t>\tau_g$ with horizon radius $R_m=2m(t)$, so that $m(\tau_g)=m_g$ , $\nu(\tau_g)=\nu_g$. Then the average distance between the black holes is (see Fig. \ref{Ignatev1})
\begin{equation}\label{R_n}
R_N(t)=\left(\frac{3}{4\pi \nu(t)}\right)^{1/3}.
\end{equation}

Assuming that starting from this moment $\tau_g$ the number of black holes does not change, we obtain:
\begin{equation}\label{N}
\nu(t)=\nu_g\left(\frac{a(\tau_g)}{a(t)}\right)^3\quad (=\nu_g e^{-3H_0(t-\tau_g)}) .
\end{equation}
Thus, from \eqref{R_n} (see Fig. \ref{Ignatev4}) we find
\begin{equation}\label{R_N}
R_N(t)=\frac{a(t)}{a(\tau_g)}\left(\frac{3}{4\pi\nu_g}\right)^{1/3}\equiv \frac{a (t)}{a(\tau_g)}R_N(\tau_g).
\end{equation}
\Fig{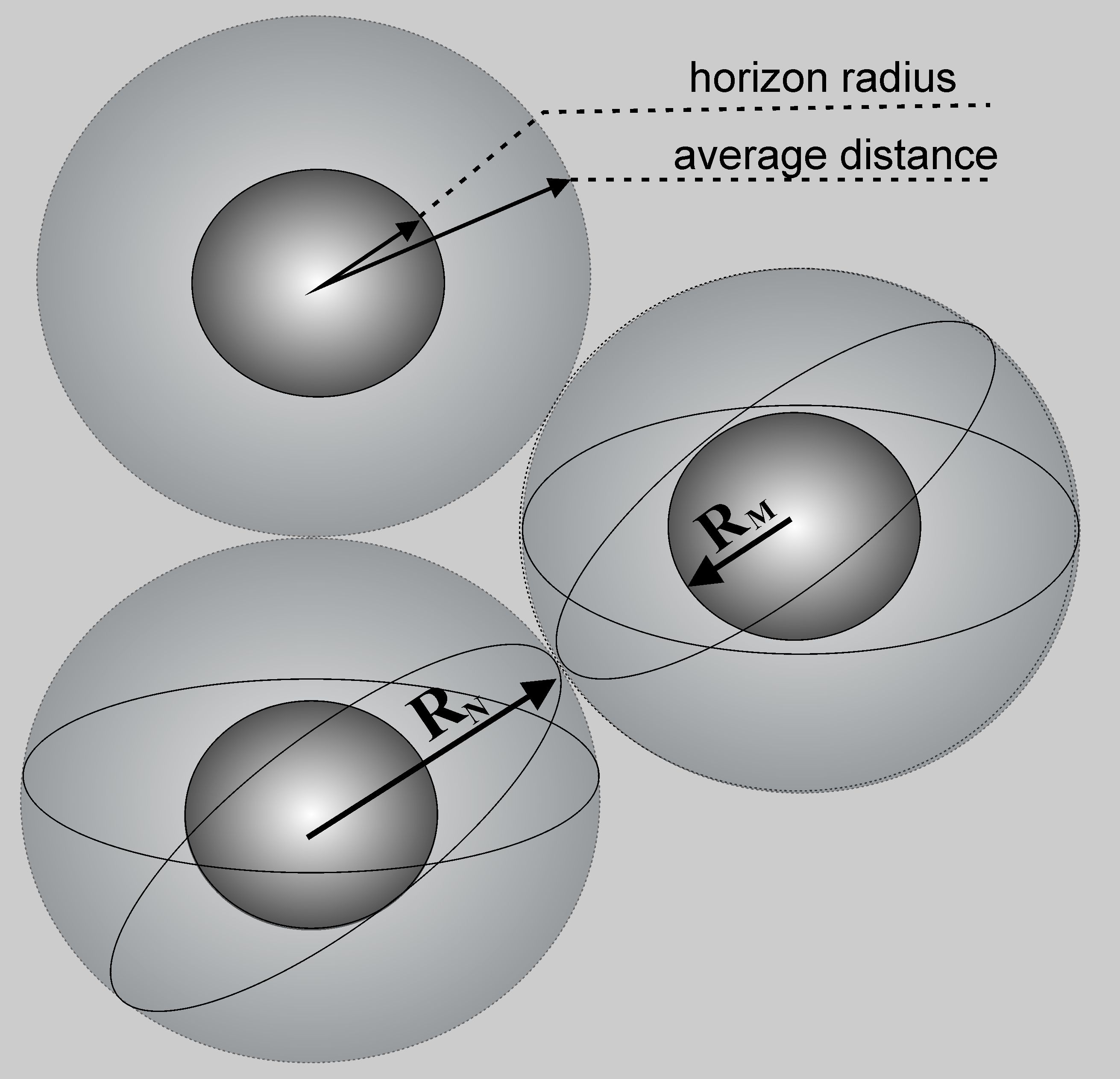}{8}{\label{Ignatev4}SSBH number density and their radii.}

Next, to estimate the radius of the black hole horizon at the inflation stage, we use the qualitative formula of the article \cite{Yu_GC_23_No4} \eqref{M_n-H0}:
\begin{eqnarray}\label{R_M}
R_M(t)=2m(t)\backsimeq 8\pi H_0^2 \frac{a^3(t)}{n^3}\nonumber\\
\backsimeq \frac{1}{\sqrt{8\pi}H_0}\mathrm{e}^{3H_0(t-\tau_g)}.
\end{eqnarray}

When the horizon radii of two black holes come into contact, the process of increasing their mass must be stopped, at least before the possible merger of the black hole. According to \eqref{R_N} and \eqref{R_M} this point in time, $\tau_{max}$, is determined by the relation
\begin{eqnarray}\label{tau_max}
\frac{a(t)}{a(\tau_g)}\gtrsim \sqrt{\delta},\Rightarrow \tau_{max}\backsimeq \frac{1}{2H_0}\ln(\delta+4),
\end{eqnarray}
where
\[\delta\equiv \frac{R_N(\tau_g)}{R_M(\tau_g)}=\sqrt{8\pi}H_0 R_N(\tau_g)>1\]
-- the ratio of the average distance between black holes to the radius of the black hole horizon at the moment of their formation. The BH density introduced above at the moment of their birth, $\nu_g$, is related to the dimensionless parameter $\delta>1$ by the relation
\begin{eqnarray}\label{nu(t)}
\nu_g=6\sqrt{8\pi}\left(\frac{H_0}{\delta}\right)^3 .
\end{eqnarray}

From \eqref{R_M} and \eqref{tau_max} we obtain an estimate for the maximum mass of SSBH
\begin{equation}\label{M_max}
M_{max}\gtrsim \frac{\delta^{3/2}}{\sqrt{8\pi} H_0}\equiv \sqrt{\frac{3}{8\pi \Lambda}}\delta^{ 3/2}.
\end{equation}
Thus, firstly, the maximum mass of a black hole is determined by only two parameters - the cosmolo\-gical constant and the density of the number of newborn black holes \eqref{nu(t)}, and, secondly, it increases with a decrease in the cosmological constant. For example, the minimum SSBH mass threshold \eqref{M_nc} $m=10^{44}$ in the case of $\delta=10^8$ is achieved at a value of the cosmological constant $\Lambda\backsimeq 10^{-65}$. Note, firstly, that according to \eqref{tau_max} the time to reach the maximum mass \eqref{M_max} is greater than or on the order of the cosmological time $\tau_0\backsimeq H^{-1}_0$ at this stage of expansion. The values of the Hubble constant in the early stages of expansion should be greater than its modern value. Secondly, as we noted, the estimating formula \eqref{R_M} gives a greatly underestimated rate of growth of the BH mass (see note \ref{mark1}), so we must make appropriate corrections to the formula \eqref{M_max}.

Summarizing the results of this section, we note that the macroscopic geometric factor is, apparently, the main one in determining the maximum para\-meters of SSBH formed in the process of cosmological evolution.

\section{Cosmological constant after\newline SSBH formation}
\subsection{Scalar field in the neighborhood of\newline SSBH}
So, since as a result of the scalar-gravitational instability of the cosmological medium of scalarly charged fermions, scalarly charged black holes are apparently formed, it is necessary, firstly, to consider the question
about such isolated static black holes. For the first time, the metric of a scalarly charged black hole in the case of a massless canonical scalar field was found in the work of I.Z. Fisher (1948) \cite{Fisher}.
In the work \cite{bronnik_fabris}, (see also reviews \cite{bronnik_rus} -- \cite{bronnik_eng}) the question of the geometry of a scalarly charged black hole with a massless scalar field ($V(\Phi)\equiv0$) was studied in detail in the absence of ordinary matter. In the classic work \cite{no-no}, the ``no-hair'' theorem was proved about the absence of scalar hair in black holes. According to this theorem, outside the black hole horizon c the scalar field can only be constant: $\Phi=\mathrm{Const}$. It should be noted that the conditions for the validity of the theorem are, firstly, the Euclidean nature of the metric at infinity, and, secondly, the presence of the event horizon itself.

Let's consider a static spherically symmetric metric in curvature coordinates (see, for example, \cite{Land_Field})
\begin{equation}\label{metric_stat}
ds^2=\mathrm{e}^{\nu(r)}dt^2-\mathrm{e}^{\lambda(r)}dr^2-r^2 d\Omega^2,
\end{equation}
in which the scalar Higgs field equation $\Phi(r)$ has the form:
\begin{eqnarray}\label{Eq_C}
\!\!\!\displaystyle \frac{1}{r^2}\frac{d}{dr}\left(r^2\mathrm{e}^{\frac{\nu-\lambda}{2 }}\frac{d}{dr}\Phi\right)-\mathrm{e}^{\frac{\nu-\lambda}{2}}\Phi(m^2_s-\alpha\Phi^2) =0.
\end{eqnarray}

In the work \cite{Yu_Scalar} the equation \eqref{Eq_C} was solved in the spatially flat metric $\nu=\lambda=0$ for the central point scalar charge $e$. For the self-action constant $\alpha=0$, the equation
\eqref{Eq_C} reduces to the well-known Yukawa equation
\begin{eqnarray}\label{EqYukava}
\displaystyle \frac{1}{r^2}\frac{d}{dr}\left(r^2\frac{d}{dr}\Phi\right)-m_s^2\Phi=0
\end{eqnarray}
and has as its solution the well-known Yukawa potential
\begin{equation}\label{Yukava}
\Phi=\displaystyle{\frac{2G}{r}\mathrm{e}^{-m_sr}},
\end{equation}
where $G$ is a scalar charge.

The self-action constant factor $\alpha\not\equiv0$ funda\-men\-tally changes the nature of solutions to the equation \eqref{Eq_C}. Now this equation has no stable solutions with zero asymptotics at infinity
\begin{equation}\label{Phi(8)=0}
\left.\Phi(r)\right|_{r\to\infty}\to 0.
\end{equation}
Stable solutions of the equation \eqref{Eq_C} in a spatially flat metric are solutions with non-zero asymptotic behavior at infinity corresponding to special stable points of the dynamical system, --
\begin{equation}\label{Phi(8)=Phi0}
\left.\Phi(r)\right|_{r\to\infty}\to \Phi_\pm=\pm \frac{m_s}{\sqrt{\alpha}}.
\end{equation}

For solutions close to stable, assuming
\begin{equation}\label{Phi0+psi}
\Phi(r)=\Phi_\pm+\phi(r),\quad (\phi\ll 1),
\end{equation}
in the linear approximation we obtain the equation instead of \eqref{EqYukava}
\begin{equation}\label{Eq_phi}
\frac{1}{r^2}\frac{d}{dr}\biggl(r^2\frac{d\phi}{dr}\biggr)+2m_s^2\phi=0.
\end{equation}
Let us pay attention to the change in sign of the massive term compared to the Yukawa equation \eqref{EqYukava}, due to which the stable solution of the equation for the Higgs field will be \cite{Yu_Scalar}
\begin{equation}\label{Phi+phi=}
\Phi(r)=\pm\frac{m_s}{\sqrt{\alpha}}+\frac{2G}{r}\cos(\sqrt{2}m_sr).
\end{equation}

The presence of a \emph{fundamental} scalar field with the Higgs potential fundamentally changes the phy\-sical picture. Now the vacuum state corresponds to one of the stable points of the Higgs potential \eqref{Phi(8)=Phi0}, which, in turn, corresponds to the zero potential energy of the scalar field
\begin{equation}\label{V(f)=0}
V(\Phi_\pm)=0.
\end{equation}
Taking into account the above, we study the solution to the complete problem of a self-gravitating scalar Higgs field. Nontrivial combinations of Einstein's equations with a cosmological constant in the metric \eqref{metric_stat}\footnote{These are combinations of the equations $^1_1$, $^4_4$ and the scalar field equation.} can be reduced to the form:
\begin{eqnarray}\label{Eq_A}
2r\Phi'^2+(\lambda+\nu)'=0;\\
\label{Eq_B}
\!\!\!\!\mathrm{e}^\lambda-1-r\nu'-\!r^2\mathrm{e}^\lambda \left[\Lambda-\frac{\alpha}{ 2}\left(\Phi^2-\frac{m^2_s}{\alpha}\right)^2\right]=0.
\end{eqnarray}

We will look for solutions to the system of equations \eqref{Eq_C}, \eqref{Eq_A}, \eqref{Eq_B} that are close to stable, assuming \eqref{Phi0+psi}. Then, in the zero approximation, due to the smallness of $\phi(r)$, the equation \eqref{Eq_C} becomes an identity, and the equation \eqref{Eq_A} gives
\begin{equation}\label{lambda=-nu}
\lambda=-\nu.
\end{equation}
As a result, the equation \eqref{Eq_B} will be reduced to the closed equation for $\nu$ (or $\lambda$)
\begin{eqnarray}
\label{Eq_B0}
r\nu'+1+\mathrm{e}^{-\nu}(1-\Lambda r^2)=0,
\end{eqnarray}
solving which, we find in the zero approximation:
\begin{equation}\label{nu0}
\nu_0=-\lambda_0=\ln\left(1-\frac{2m}{r}-\frac{\Lambda r^2}{3}\right),
\end{equation}
where $m$ is the constant of integration. Thus, in the zeroth approximation we obtain the well-known Schwarzschild-de Sitter solution \cite{Edd}:
\begin{eqnarray}\label{Shvarc-deSit}
ds^2=&\displaystyle \left(1-\frac{2m}{r}-\frac{\Lambda r^2}{3}\right)dt^2 &\nonumber\\
& \displaystyle -\frac{dr^2}{\displaystyle 1-\frac{2m}{r}-\frac{\Lambda r^2}{3}}-r^2d\Omega^2.&
\end{eqnarray}

Due to \eqref{Eq_A}, in the first approximation of the smallness of $\phi$ the relation \eqref{lambda=-nu} is preserved, and therefore the equation \eqref{Eq_B0} is also preserved.
It follows that\emph{ the metric \eqref{Shvarc-deSit} is preserved in the approximation linear in $\phi$}. Therefore, in a linear approximation, the field equation \eqref{Eq_C} can be consi\-dered against the background of the Schwarzschild - de Sitter solution \eqref{Shvarc-deSit}:
\begin{eqnarray}\label{Eq_C1}
\displaystyle \frac{1}{r^2}\frac{d}{dr}\left(r^2\mathrm{e}^{\nu_0}\frac{d}{dr}\phi\right)+ 2m^2_s\mathrm{e}^{\nu_0}\phi=0.
\end{eqnarray}

Without posing in this article the problem of finding solutions to the equation \eqref{Eq_C1} and studying their behavior near the horizons of the metric \eqref{Shvarc-deSit}, we only note that, in general, the solution is in the form \eqref{Phi0+psi}
with $\Phi(\infty)=\Phi_\pm=\mathrm{Const}$ \eqref{Phi(8)=Phi0} does not contradict the theorem about the absence of scalar hairs in black holes. At the same time, the Higgs potential is implicitly included in the solution of field equations, firstly, through the vacuum value of the scalar potential \eqref{Phi(8)=Phi0} and, secondly, through the renormalization of the bare cosmological constant $\Lambda_0$ (see. , for example, \cite{Yu_GC_23_No4}), decreasing its value.
\begin{equation}\label{lambda0->Lambda}
\Lambda=\Lambda_0-\frac{1}{4}\frac{m_s^4}{\alpha}.
\end{equation}
\subsection{Cosmological factor}
It must be remembered, however, that the properties of a scalarly charged black hole discussed above refer to an isolated static system. In fact, the process of the formation of a black hole occurs in the cosmological environment, which determines its evolution. In this regard, let us again turn to Fig. \ref{Ignatev1} -- \ref{Ignatev2}, illustrating the evolution of the cosmological medium during the formation of a black hole for the standard model $\mathbf{SM}$ \eqref{ParSM}.

In this regard, the question arises of how to combine the model of an isolated static scalarly charged BH with the model of a BH in a cosmological environment. In particular, what can happen to the quasi-vacuum state of this black hole, corresponding in this case to the value of the scalar potential $\Phi_\pm=1$, which in the cosmological environment after the transition should tend to zero, thereby violating the quasi-vacuum nature of the state and causing instability of the scalar field . This rather serious issue requires additional research, which we hope to conduct in the future.

For now, we will explore the essence of this issue from the point of view of the qualitative theory of ordinary differential equations (see, for example, \cite{Bogoyav}).
To do this, consider the equation of the scalar field $\Phi(x,t)$ with potential energy $V(\Phi)$ in flat space-time:
\begin{equation}\label{EqPhi_TX}
\ddot{\Phi}-\Phi''+V'_\Phi(\Phi)=0,
\end{equation}
where, as usual, the dot denotes derivatives with respect to $t$, and the prime -= with respect to $x$. Let's consider two fundamentally different situations - the case when $\Phi=\Phi(t)$, and the case when $\Phi=\Phi(x)$. We will call the first situation the T-situation, and the second the X-situation. The corresponding field equations are obtained from \eqref{EqPhi_TX}:
\begin{eqnarray}\label{EqPhi_T}
\mathbf{T}:\; & \ddot{\Phi}+V'_\Phi(\Phi)=0;\\
\label{EqPhi_X}
\mathbf{X}:\; & \Phi''-V'_\Phi(\Phi)=0.
\end{eqnarray}
According to the qualitative theory of differential equations, in both situations the singular points $\Phi_i$ of the corresponding normal system of differential equations are determined by the equations:
\[V'_\Phi(\Phi_i)=0,\]
and the eigenvalues of the characteristic matrix of the system $\lambda_s$, in turn, are determined through these singular points:
\begin{eqnarray}
\mathbf{T}:\; & \lambda_i=\pm\sqrt{-V''_{\Phi\Phi}(\Phi_i)}=0;\nonumber\\
\mathbf{X}:\; & \lambda_i=\pm\sqrt{V''_{\Phi\Phi}(\Phi_i)}=0.\nonumber
\end{eqnarray}
In the case of the Higgs potential
\[V(\Phi)=-\frac{\alpha}{4}\left(\Phi^2-\frac{m^2_s}{\alpha}\right)^2\]
there are singular points of the system
\[\Phi_0=0,\quad \Phi_\pm=\pm\frac{m_s}{\sqrt{\alpha}}.\]
So, we get for the eigenvalues:
\begin{eqnarray}
\mathbf{T}:\; & \lambda_0=\pm im_s; & \lambda_\pm=\pm\sqrt{2}m_s;\nonumber\\
\mathbf{X}:\; & \lambda_0=\pm m_s; & \lambda_\pm=\pm i\sqrt{2}m_s=0.\nonumber
\end{eqnarray}

Thus, according to the qualitative theory of differential equations in the T-situation, the zero singular point $\Phi_0=0$ is an attracting focus (cycle), and the singular points $\Phi_\pm$ are saddle points, i.e., unstable points of the system. In the case of the X-situation, on the contrary, the zero singular point $\Phi_0=0$ is a saddle point, i.e., unstable, and the singular points $\Phi_\pm$ are attracting foci (cycles). This explains why in the static case the scalar potential tends to a quasi-vacuum solution at infinity \eqref{Phi(8)=Phi0} (X-situation), which is unstable for the cosmological T-situation, while stable for the cosmological situation at $ t\to\infty$ the solution is zero. The collision of these systems, which are opposite from the point of view of differential equations, is the problem that we have to solve in the future.

One can, of course, object to this conclusion. Indeed, in the standard model of a scalar field with a parabolic potential, the zero singular point is preserved, as are the eigenvalues of the characteristic matrix at this point. But then it turns out that the solution $\Phi=C_1\exp(-m_s x)$ at this point $\Phi\to0$ for $x\to\infty$ is not stable for a static field, just like the Yukawa solution. Yes, indeed, these solutions, strictly speaking, are not stable, since they do not have a growing branch in comparison with general solutions. The damped (only for $x\to+\infty$!) solution corresponds to particular initial conditions
\[ \Phi(0)=\Phi_0,\; \Phi'(0)=-\frac{\Phi_0}{m_s}.\]
With a slight variation of these conditions, a second branch appears in the solution, diverging expo\-nen\-tially quickly at positive infinity, which corresponds to classical Lyapunov instability. A completely si\-milar situation arises with the Yukawa potential in the case of spherical symmetry. In the case of a parabolic potential of a scalar field, this contra\-dic\-tion can be quite simply eliminated, using the fact that there is only one singular point of the X- and T-systems. If there are several singular points with different characteristics, this contra\-dic\-tion cannot be easily eliminated. Although even in this case there is a way out, as long as the cosmological system is, albeit in an unstable, but rather long inflation phase, as in Fig. \ref{Ignatev2} -- \ref{Ignatev3}. In this case, the cosmologically unstable state (T-situation) with $\Phi=1$ is at the same time stable for the X-situation, i.e., for a scalarly charged BH.

If we assume that the general cosmological tendency in the course of evolution nevertheless turns out to be dominant, then we must accept as a fact that the scalar field in the outer regions close to the horizons of the BH should completely disappear during the course of cosmological evolution. Thus, there should be a strong drop in the value of the cosmological constant, possibly even to zero, in accordance with the formula \eqref{lambda0->Lambda}. In this case, under the horizon, the scalar field can remain in a state close to stable $\Phi=\Phi_\pm$ \eqref{Phi(8)=Phi0}.

\subsection{Macroscopic picture\newline of the Universe with black holes}
Let's find out what the situation can lead to when the Universe is filled with supermassive scalarly charged black holes, surrounded by fermionic matter in the absence (or in the presence of a weak scalar field).
In the early works of the Author \cite{YuI_GTO14}, \cite{YuI_GC07} \cite{YuI_GTO20} the foundations of the statistical theory of relativistic classical systems with gravitational inte\-rac\-tion were formulated. In the articles \cite{YuI_Pop_IzvVuz}, \cite{YuI_Pop_ASS}, based on this theory, a kinetic equation for massless particles in the macroscopic Friedmann world was derived, taking into account the gravitational inte\-rac\-tion with microscopic spherical symmetric local fluctuations of the metric generated by point sources of mass. In this case, the cosmological evolution of spherical local fluc\-tua\-tions generated by point masses was taken into account, studied in \cite{YuPhysA} for the ultrarelativistic equation of state of matter in the Universe, and in \cite{Moroca1} and other works - for the equation of state of an ideal fluid with an arbitrary constant barotropic coefficient . Finally, in \cite{Yu_BH_19} these results were applied to obtain an effective energy-momentum correlation tensor for quadratic fluctuations of the gravitational field of black holes (BHs), arising due to the overlap of their gravitational attraction regions in the macro\-sco\-pi\-cally spatially flat Friedmann Universe. The resulting expression for the EI components at the nonrelativistic stage of evolution of the material component of the cosmological environment, adapted to our notation, has the form:

\begin{eqnarray}\label{dT}
\delta T^i_k=-\frac{1}{8\pi}\delta \overline{G^{(2)}\!\ ^i_k}\nonumber\\
=3\pi \nu_sr_s\left(\frac{2m(t)}{r_s}\right)^2 \delta^i_k\equiv \langle\varepsilon_g\rangle\delta^i_k,
\end{eqnarray}
where $\langle\varepsilon_g\rangle$ is the average correlation energy density, $\nu_s=\nu(r_s)$ is the average BH density per accom\-panying volume, $\mu(t)=m(t)/a(t )$ is the reduced mass of the black hole, $r_s$ is the sound horizon at the moment of transition to the total non-relativistic state of cosmological matter. Assuming that after the completion of the process of exponentially rapid mass growth, the matter becomes non-relativistic and the SSBH mass remains practically unchanged (only due to slow gas accretion), we obtain:
\begin{equation}\label{<e>}
\langle\varepsilon_g\rangle\sim 12\pi\frac{\nu_s m^2_{ssbh}}{r_s}= \mathrm{Const}.
\end{equation}
Thus, at the nonrelativistic stage of expansion after completion of the SSBH formation, the Universe can be described by a cosmological model with an effective cosmological constant
\[\Lambda_{eff}=\langle\varepsilon_g\rangle, \]
generated by quadratic correlations of local gravi\-ta\-tional fields SSBH.

\section{Conclusion}
Note that the BH number density $\nu(t)$ appears in section \ref{razdel2}, therefore the formulas for the effective constant \eqref{<e>} and the maximum achievable mass of the BH \eqref{M_max} are quite strictly related to each other.
Thus, to clarify the correctness of the estimate \eqref{<e>} of the modern value of the cosmological constant, observational data on the maximum mass of SSBH $m_{max}$, their average density $\nu(t)$ and the radius of the sound horizon at the moment of matter at non-relativistic stage.

\noindent \textbf{Acknowledgements}\\
The author is grateful to the participants of the seminar of the Department of Relativity and Gravity at Kazan University for a useful discussion of some aspects of the work. The Author is especially grateful to professors S.V. Sushkov and A.B. Balakin.

\noindent \textbf{Subsidization}\\
The work was carried out using subsidies allocated as part of state support for the Kazan (Volga Region) Federal University in order to increase its competitiveness among the world's leading scientific and educational centers.
\setcounter{section}{0}
\setcounter{equation}{0}
\setcounter{figure}{0}


\end{document}